\documentclass[journal=jacsat,manuscript=article]{achemso}

\usepackage{latexsym,graphicx,graphics}
\usepackage{appendix}
\usepackage{amsmath,amssymb,epsfig}
\usepackage[utf8]{inputenc}
\usepackage[mathscr]{euscript}
\usepackage{float}
\newcommand{\abs}[1]{\left| #1 \right|}

\newcommand{\bb}[1]{ \mbox{\boldmath$ #1$}}

\newcommand{\rv}{\bb r}

\newcommand{\Eq}[1]{Eq.~(\ref{#1})}
\newcommand{\Eqs}[2]{Eqs.~(\ref{#1})--(\ref{#2})}

\newcommand{\unit}[1]{\bb{\hat{#1}}}

\title{Dark Mode - Faraday Rotation Synergy for Enhanced Magneto-Optics}

\author{Y. Mazor}
\author{M. Meir}
\author{Ben Z. Steinberg}
\email{steinber@eng.tau.ac.il}
\affiliation{
School of Electrical Engineering, Tel-Aviv University, Ramat Aviv, Tel-Aviv 69978 Israel
}%

\date{\today}
\begin{document}

\begin{abstract}
We examine the efficacy of Dark-mode plasmonics as a platform for enhanced magneto-optics. Dark-mode of a small particle consists of two co-existing equal-intensity and mutually opposing dipolar excitations.  Each of these two opposing dipoles may even resonate at or near the dark-mode frequency, but the net dipole moment vanishes due to the mutual cancelation between the opposing dipoles. We show that application of external magnetic bias may alleviate the intense destructive interference. Furthermore,
under external magnetic bias the opposing dark-resonances of a plasmonic particle shift in opposite directions and create a region of extremely sensitive Faraday rotation. We show that the magnetized dark resonance in lossless Ag-like particle may provide more than 20 degrees rotation under magnetic fields of the order of 1-2 Tesla, exhibiting magneto-plasmonic activity that is 2-3 orders of magnitude larger than that observed in conventional plasmonic particle of the same material.\newline
\end{abstract}

\maketitle

%\section{Introduction}

%\section{Introduction}
Dark modes of an open optical structure can be described as states of excitations that incorporate mutually opposing local dipoles whose far-fields interfere destructively. The net dipolar excitation then vanishes, resulting in a significant reduction of the far-field radiation, and consequently a reduction of the associated radiation damping and bandwidth \cite{Stockman,BoundStatesPhotonics,BrightDark,Maier1,DMExcitationLocalizedEmitter,DarkPlasmonics,DarkAlu,FanoHalas}. This, in turn, may trap optical fields in a structure that is inherently coupled to a continuum. More formally, dark modes can be viewed as manifestations of discrete eigenvalues embedded within the continuous spectrum of the associated non-compact scattering operator.
Dark modes were suggested as candidates for electromagnetic energy storage, enhanced biological and chemical sensors, and nanoscale waveguides. These modes can be supported by simple structures such as nano-dimers (see, e.g.~the ``anti-bonding'' plasmons in Ref. [\citenum{Stockman}]), trimers\cite{DarkPlasmonics}, clustered nano-rods\cite{BrightDark} and spheres\cite{DarkAlu,FanoHalas}.

In a seemingly unrelated research endeavor, nonreciprocal magneto-optics and its implementation for one-way waveguides, optical isolators and circulators, and Faraday rotators, have been under intensive study \cite{Fan,HadadSteinberg,MazorSteinbergFlat,OneWayStrips,Wire,Circulator}.
Currently the major drawback of nonreciprocal magneto-optics is the requirement for strong magnetic bias $\bb{B}_0$. Efforts to reduce $\bb{B}_0$ for various applications (e.g.~Faraday polarization rotation) in plasmonic structures can be found, e.g.~in \cite{GoldCoatedIronOxide,Graphene1, GrapheneACS}. The work in \cite{GoldCoatedIronOxide} reports on an experimental evidence for a 2-3 fold enhancement of magneto-optical activity in coated nano-particles. The efforts in \cite{Graphene1, GrapheneACS} are limited to graphene metasurfaces.

\begin{figure}[h]
\begin{center}
\noindent
  \includegraphics[width=8cm]{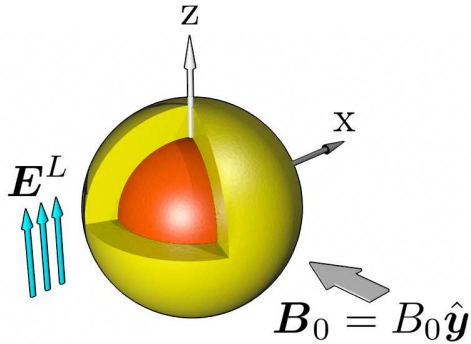}
  \caption{The particle geometry. The shell and the core are made of Drude-metals with plasma frequencies $\omega_{p1,2}$, and have radii $a_{1,2}$, respectively.}
  \label{fig1}
\end{center}
\end{figure}

Here we study the effect of bias magnetization on plasmonic particle dark modes, and explore its potential applications as a new platform for non-reciprocal optics. As a simple and physically transparent test-case, we consider the core-shell spherical particle shown in Fig.~\ref{fig1}, made of two plasmonic materials with close, but not identical, plasma frequencies. The structure is excited by a linearly polarized local field $\bb{E}^L(\rv)=\unit{z}E^Le^{iky}$. When properly designed, the linear dipole in the shell ($\bb{p}_1=\unit{z}p_1$) and the core ($\bb{p}_2=\unit{z}p_2$) resonate and possess equal magnitudes and opposite phases, thus creating a resonating linearly-polarized dark mode.

\begin{figure}[h]
\begin{center}
\noindent
  \includegraphics[width=10cm]{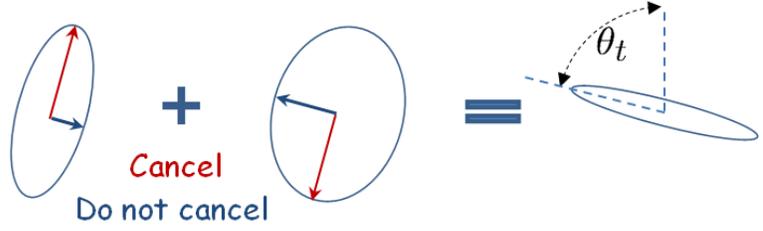}
  \caption{Illustration of the physical mechanism which generates the large polarization rotation angles. Each of the polarization ellipses represent the dipole response inside one component of our core-shell structure. They are only slightly rotated from the vertical exciting field. However, in the chosen operation regime, the principal axes nearly cancel each other, where the minor axes do not (this is possible since the ellipses have different principle axes) resulting in a net polarization ellipse that is significantly rotated.}
  \label{fig0}
\end{center}
\end{figure}

Under magnetic bias $\bb{B}_0=\unit{y}B_0$ this dark resonance splits into two separate resonances corresponding to elliptically polarized dipoles in the core and the shell. The ratio between the ellipse's principal axes in each domain is determined by $\omega_b/\omega_{p\,1,2}$ where $\omega_b=-qB_0/m_e$ is the cyclotron frequency. Since $\omega_{p\,1}\ne\omega_{p\, 2}$ the ellipses cannot be identical. They are of nearly destructive interference (dark mode) along the major axis ($\unit{z}$) and non-destructive along $\unit{x}$. Hence the net dipole rotates at large angles already at very low $B_0$. The principle of this mechanism is illustrated in Fig.~\ref{fig0}.

Finally, we note that the effect of material loss is neglected here. The efforts to combat loss span a wide range of approaches and methods, either passive \cite{Khurgin_Elusive, PRX, Naik2, MaradudinBook, PendrySpoof, SpoofExperimental, SpoofNonRecip} or active \cite{Active_Book, Capolino1, Capolino2}. In the former the use of doping or alloying techniques has lead to new plasmonic materials with $\epsilon<0$ at the IR and optical ranges, with loss that is five times lower than that of Ag, as reported theoretically and experimentally in [\citenum{PRX, Naik2}]. Other methods, including e.g. band-engineering or use of TCO's are under current investigation - for an overview see Chapt.~6 in Ref. [\citenum{MaradudinBook}]. A somewhat different approach (but still passive) is the ``spoof-plasmon'' idea \cite{PendrySpoof, SpoofExperimental, SpoofNonRecip} in which one mimics a plasmonic wave and $\epsilon<0$ in patterned metals at much lower frequencies, where the metal loss is practically zero. This approach has been applied also for non-reciprocal devices [\citenum{SpoofNonRecip}]. Using active approach, complete loss compensation by gain using spasers is discussed e.g. in Chapt.~1 in Ref. [\citenum{Active_Book}]. Core-shell particles of metallic shell and dielectric core with embedded fluorescent dyes for gain are studied in [\citenum{Capolino1}]. A reversed geometry with loss compensating gain is studied in [\citenum{Capolino2}]. To contrast, we stress that the magnetic fields required to achieve a useful non-reciprocity in nano photonics are many orders of magnitude stronger than any practical level, and this fact is \emph{independent of material loss}. As we show below, the Faraday rotation in a conventional plasmonic particle, achieved by field strengths of 1.5T, is a fraction of a degree even if the particle is made of lossless Drude-like metal. Hence, the efforts described above to combat loss, as successful as they prove to be, will not solve the problem of achieving practical non-reciprocity in nano-particles. This problem deserves a research endeavor by its own; this goal is addressed here.

%\section{Formulation}
\section{Formulation}
Consider the core-shell spherical plasmonic particle shown in Fig.~\ref{fig1}. The shell (core) is made of Drude-metals with plasma frequency $\omega_{p1}$ and tensor electric susceptibility $\bb{\chi}_1$ ($\omega_{p2}$ and $\bb{\chi}_2$).
The outer shell radius and core radius are $a_1$ and $a_2$, respectively. The particle is electrically small hence we use polarizability theory and discrete dipole approximation to model the particle's dynamics. The particle response to an exciting electromagnetic field is fully characterized by the dipole response $\bb{p}=\bb{\alpha}\bb{E}^L$, where $\bb{E}^L$ is the local electric field and $\bb{\alpha}$ is the particle's \emph{dynamic} polarizability tensor. It can be written as
\begin{equation}\label{eq1}
\bb{\alpha}^{-1}=\bb{\alpha}_h^{-1}-\frac{ik^3}{6\pi\epsilon_0}\bb{I}
\end{equation}
where $\bb{\alpha}_h$ is the \emph{static} polarizability, and the imaginary term on the right represents radiation damping \cite{RadiativeCorrection}. We derive $\bb{\alpha}_h$ by solving the Laplace equation under $\unit{z}$-directed local electric field $E^L$, with the appropriate boundary conditions. For non-magnetized particle the problem reduces to a core-shell spherical structure with scalar permittivities $\epsilon_{2,1}$. Analytic expressions for the corresponding $\bb{\alpha}_h$ can be found, e.g.~in \cite{SihvolaBook}. Isotropic core-shell particles were suggested also for a variety of applications, including compact waveguides \cite{AluEngheta1,AluEngheta2}, non-linear switches \cite{Cloaks}, absorbers \cite{TretyakovCoreShell}, and more. However, under magnetization $\bb{\chi}_{1,2}$ become tensors. To the best of our knowledge, a study of the  polarizability and dipole-responses of the entire structure and of the shell and core regions for tensor $\bb{\chi}_{1,2}$ are not currently available in the literature. See Supporting Information (SI) for analytical derivation of the fields, core and shell polarization, and polarizability of the magnetized core-shell particle. In the SI and below we also compare the results of this analysis to exact full-wave simulations using the CST$^\copyright$ commercial software.

%\section{Results}

\section{Results}
\begin{figure}[h]
\begin{center}
\noindent
\hspace*{-0.33in}
  \includegraphics[width=12cm]{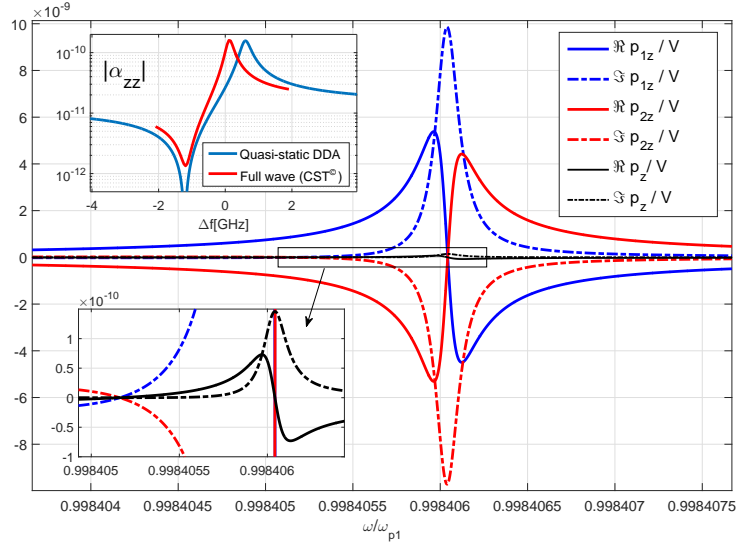}
  \caption{The dark resonance dipole excitations, normalized to the particle volume. $\bb{p}_{1,2}$ are the shell and core dipoles, respectively, and $\bb{p}=\bb{p}_1+\bb{p}_2$. The net dipole of the sphere identically vanishes at the ``dark-point'' - a frequency located on the left side close to the resonance. The inset in the lower left side shows a blow-up picture near the resonance point. The inset in the upper left side compares the quasistatic analysis of $|\alpha_{zz}|$ to that obtained by full wave analysis with a commercial software. The dark point is clearly seen left of the dark resonance.}
  \label{fig2}
\end{center}
\end{figure}

Using the formulation developed in the Supporting Information (SI), we have extracted the shell and core dipoles, net dipole and polarizabilites for $a_1/a_2=3$, and for $\omega_{p1}/\omega_{p2}=1.005$. We note that the latter is a convenient representative of the typical range achieved with various metals. For example, for Au-Ag structure it is about 1.0022 according to \cite{PFreq1}, or 1.00137 according to \cite{PFreq2}. Different ratios (smaller and larger) are obtained by using other metals as can be seen from the data provided there. Furthermore, one may tune the plasma frequency by using the techniques reported e.g. in references \cite{Khurgin_Elusive, PRX, Naik2, MaradudinBook, PendrySpoof, SpoofExperimental, SpoofNonRecip, Active_Book}. We chose the ratio 1.005 just for convenience. The incident field strength is $\bb{E}^L=\unit{z}1$v/m.
Under these parameters the particle possesses two resonances (see SI). The lowest one is essentially similar to the resonance of a uniform sphere in the sense that the dipolar moments of the shell and core are in-phase, and the associated polarization densities are very similar. In contrast, the upper one resembles a dark-mode nature - the dipolar moments are in opposite phases. We concentrate here on the upper resonance.
Figure \ref{fig2} shows the shell and core dipoles $\bb{p}_{1,2}=\int_{v_{1,2}} \bb{P}_{1,2}dv$ where $v_{1,2}$ are the shell and core volumes, and $\bb{P}_{1,2}$ are their polarization densities. Throughout the entire domain $\bb{p}_1\approx -\bb{p}_2$. The equality is exact at the ``dark point'' $\omega_d/\omega_{p1}=0.9984052$ where $\bb{\alpha}$ vanishes, and so does the net dipolar moment of the entire particle. $\bb{p}_{1,2}$ resonate at a slightly higher frequency, creating net resonance of extremely narrow line shape at $\omega_{dr}/\omega_{p1}=0.99840605$ - the ``dark resonance''. The upper inset shows $|\alpha_{zz}|$ on a logarithmic scale and compares it with exact full-wave numerical simulation done with the CST$^\copyright$ commercial software. Here the parameters are $a_1=15$nm and $\omega_{p1}\approx 13\times 10^{15}\mbox{Rad/Sec.}$ (corresponding to Ag). The dark point, the dark resonance, and the non-symmetric Fano-like line shape \cite{FanoHalas} are seen. There is an excellent agreement between the quasi-static analysis presented in the SI, and the full-wave simulations.

\begin{figure}[h]
\begin{center}
\noindent
\hspace*{-0.27in}
  \includegraphics[width=12cm]{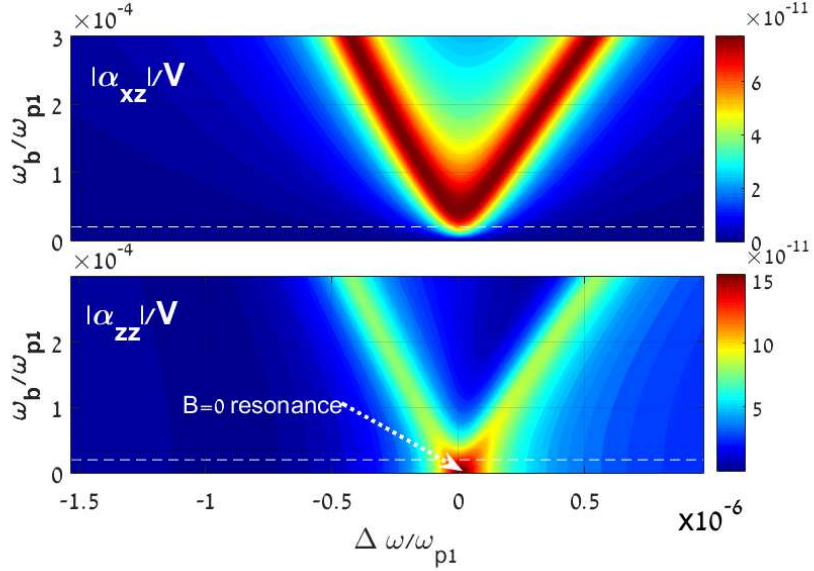}
  \caption{Absolute values of the entries of $\bb{\alpha}$ versus frequency and magnetization. The horizontal dashed line corresponds to $B_0=1.5$T for Ag-like shell. Note the resonance splitting due to bias magnetization.}
  \label{New_Fig_4}
\end{center}
\end{figure}

Figure \ref{New_Fig_4} shows the absolute components of the polarizability $\bb{\alpha}$, namely $\abs{\alpha_{xz}}$ and $\abs{\alpha_{zz}}$ as a function of frequency for a range of magnetization levels. It is seen that for $\omega_b=0$ ($B_0=0$) there is no cross-polarization: $\alpha_{xz}=0$. Cross-polarization and resonance splitting take place for $B_0\ne 0$. At low levels of $B_0$, the splitting is visually ``smeared'' due to the particle's radiation loss as can be seen in the bottom part of both panes of this figure. The horizontal dashed line corresponds to $B_0=1.5$T for Ag-like shell.

\begin{figure}[h]
\begin{center}
\noindent
\hspace*{-0.27in}
  \includegraphics[width=12cm]{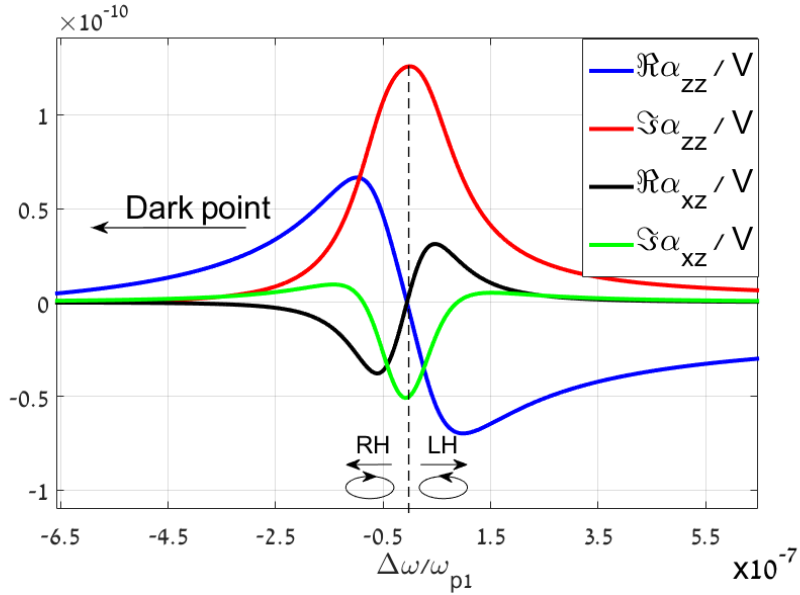}
  \caption{The polarizability at the Dark resonance regime with $\omega_b/\omega_{p1}=2\times 10^{-5}$, corresponding to the white dashed line of Fig.~\ref{New_Fig_4}. The real (imaginary) part of $\bb{\alpha}$ are odd (even) in $\omega$. Hence, the elliptical polarization changes from right-hand to left-hand across the central frequency.}
  \label{fig3}
\end{center}
\end{figure}

\begin{figure}[h]
\begin{center}
\noindent
\hspace*{-0.27in}
  \includegraphics[width=12cm]{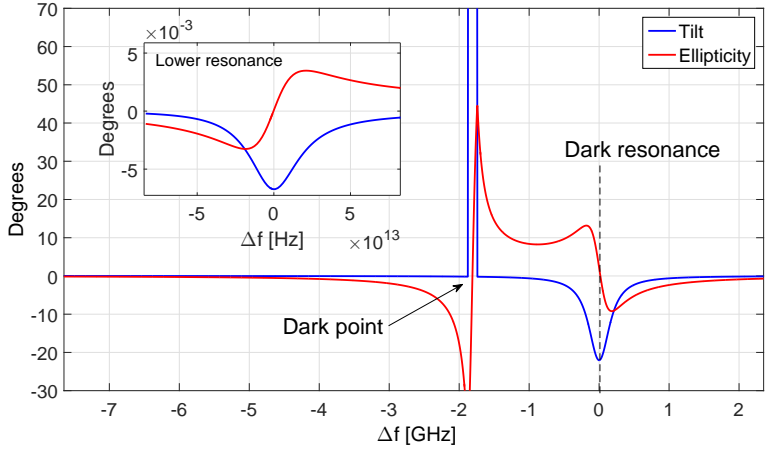}
  \caption{Dipole tilt and ellipticity angles $\psi,\theta_e$ under magnetization of 1.5T in Ag-like core-shell particle. The inset shows the lower-resonance domain. The ``dark point'' and ``dark resonance'' refer to their locations in the pre-magnetized particle, and $\Delta f$ is the shift in GHz from the central frequency of Fig.~\ref{fig2}, using $\omega_{p1}$ of Ag.}
  \label{fig4}
\end{center}
\end{figure}

Figure \ref{fig3} shows the particle dynamic polarizabilities $\alpha_{zz},\alpha_{xz}$ under magnetic bias $\unit{y}B_0$ that corresponds to $\omega_b/\omega_{p1}=2\times 10^{-5}$, corresponding to $B_0\approx 1.5$ Tesla for Ag or Au material. The combined dark resonance splits into two separate resonances. At this level of magnetization both are hidden within a single line-shape (since the line-width, due to radiation damping, is larger than the split). As seen, at the central frequency the polarizabilities real-part vanish, and one obtains a \emph{linear} polarization, rotated by the angle $\psi=\tan^{-1}(\Im \alpha_{xz}/\Im \alpha_{zz})\approx\tan^{-1}(-0.385)=-21^\circ$ relative to the exciting field. As one shifts away from this central frequency, the dipole polarization becomes elliptical, with \emph{tilted} principal axis.  In this domain, the tilt angle $\psi$ alone is not a sufficient measure of the effective polarization rotation; one needs also the \emph{ellipticity} angle $\theta_e$ defined via $\tan\theta_e=b/a$ where $a,b$ are the ellipse major and minor principle axes. $\theta_e=0^\circ$ ($\theta_e=\pm 45^\circ$) correspond to linear (circular) polarizations. We emphasize that the results in this figure are verified by exact full-wave numerical simulation using the $\mbox{CST}^\copyright$ commercial software (see SI).
Figure \ref{fig4} shows $\psi$ and $\theta_e$ of the dipole's elliptical polarization vs. frequency, for the particle parameters of Fig.~\ref{fig3}. The main figure shows $\psi,\theta_e$ for the upper resonance domain. The inset shows the lower resonance, that exhibits essentially the same dynamics of a conventional magnetized plasmonic particle. In the upper resonance region, $B_0\approx1.5$T provides dipole tilts of three orders of magnitude larger compare to the tilts in the lower resonance. We refer the reader also to Fig.~3 in the SI and the pertaining discussion there, showing $\psi,\theta_e$ for a conventional particle of lossless Ag. The maximal rotation for $B_0\approx1.5$T is less than $0.06^\circ$.


%\section{Applications}
\section{Applications}
\begin{figure}[h]
\begin{center}
\noindent
  \includegraphics[width=4.5cm]{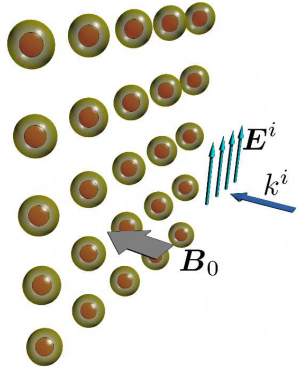}
  \caption{An array of dark mode particles for polarization rotation.}
  \label{fig5}
\end{center}
\end{figure}

Consider a metasurface made of periodic two-dimensional array of our dark-mode particles, shown schematically in Fig.~\ref{fig5}. It can be used as a Faraday rotator for applications such as optical isolators. The array of rotated dipoles would generate also a net rotated E-field, in the same manner as suggested, e.g.~in \cite{GrapheneACS}. There is a well established and accurate theory for the electrodynamic properties of such metasurfaces--see e.g.~Sec.~4.5 in \cite{TretyakovBook}--and its application to the present particles is straightforward. We note that this theory takes into account the dipole-dipole interaction in the array and its effect on the reflection and transmission (see e.g. the interaction constant $\beta$ there), as well as the individual particle's properties.

\begin{figure}[h]
\begin{center}
\noindent
%\hspace*{-0.25in}
  \includegraphics[width=12cm]{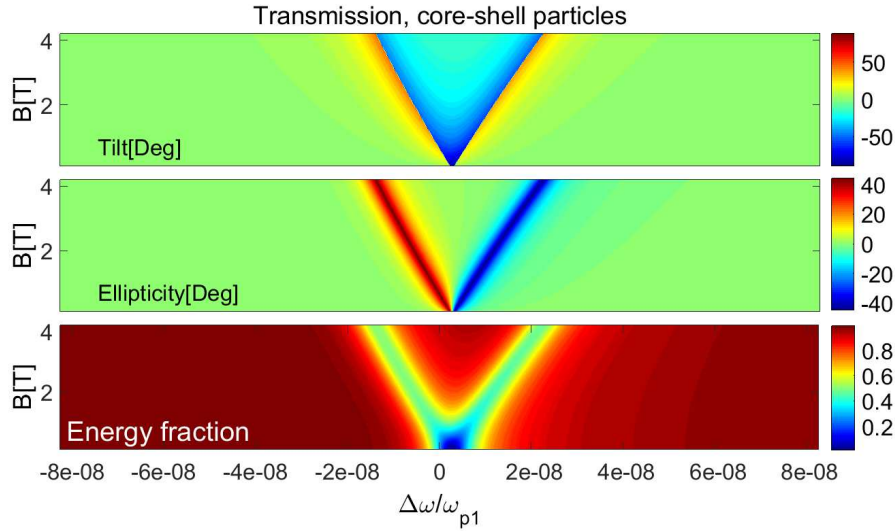}
  \caption{Transmission properties of an array of core-shell particles at the dark mode. $\Delta \omega$ is the shift from the central frequency of Fig.~\ref{fig2} (the dark resonance without magnetization), using $\omega_{p1}$ of Ag.}
  \label{fig6}
\end{center}
\end{figure}

\begin{figure}[h]
\begin{center}
\noindent
%\hspace*{-0.25in}
  \includegraphics[width=12cm]{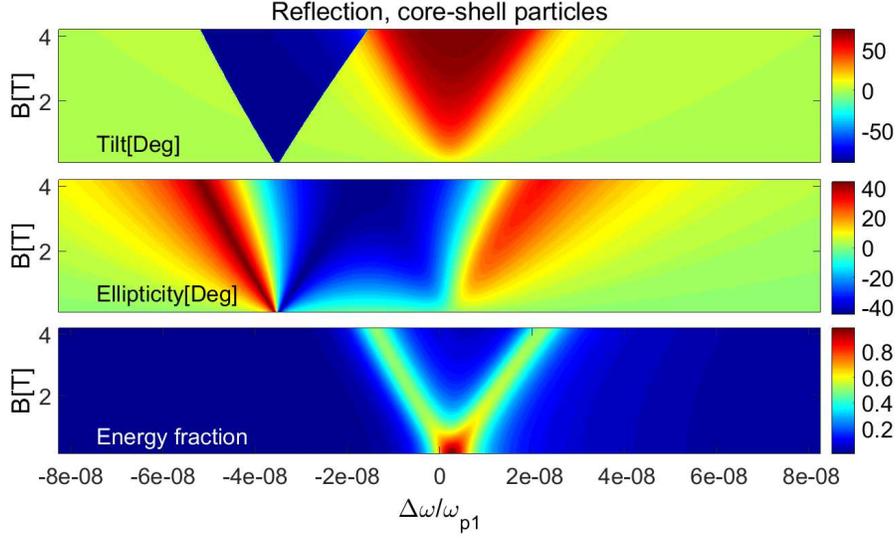}
  \caption{Reflection properties of an array of core-shell particles at the dark mode.}
  \label{fig7}
\end{center}
\end{figure}

 The far field transmission and reflection properties of this array are shown in Fig.~\ref{fig6} and \ref{fig7}, respectively, as functions of $\Delta\omega/\omega_{p1}$ and $B_0$. The parameters used are $f_{p1}=2.068\times10^{15}$Hz, $\omega_{p1}/\omega_{p2}=1.001$, $a_1=3a_2=15$nm, and $D=65$nm. For $B_0=1.5$T and near $\Delta\omega/\omega_{p1}=0$ the transmitted far field tilt angle is about $25^\circ$, with nearly linear polarization. The power transmitted into this tilted field is about 50\% of the incident field. As $B_0$ increases the transmitted field resonance splits into two branches, with large tilt and ellipticity. Essentially the same dynamics is observed for the reflected field in Fig.~\ref{fig7}. Note that for $B_0=0$ the screen resonance is slightly blue-shifted with respect to the resonance of a single unbiased particle (the latter is at $\Delta\omega=0$). This shift is due to the dipole-dipole interactions. Also note that the splitting effect shown for an individual particle in Fig.~\ref{New_Fig_4}, exists also here.

 Finally we note that the screen transmission coefficient obtained from the semi-analytical approach used here, is compared to that obtained from full-wave simulation using the CST$^\copyright$ in Fig.~4 of the SI. Since the latter is computation-intensive the comparison there is done only for the magnetic bias of 1.5Tesla and not as a map of the type shown in Figs.~\ref{fig6},\ref{fig7} here. It is seen that the results agree well.

\vspace*{0.35in}

\subsection{The effect of loss}
Clearly, material loss degrades the performance of the effect reported here. A parametric study of the dark-mode based Faraday rotation under low material loss is included in the SI. In the range of parameters tested it is seen there that although the tilt angle may be reduced significantly, it is still orders of magnitude larger than the tilt obtained by conventional particles under the same conditions.

%\section{Conclusions}

\section{Concluding remarks}
A preliminary study of a synergetic interaction between dark mode and Faraday rotation is reported. The study uses the core-shell geometry with two nearly equal plasma frequencies as a test-case in which the mathematical analysis is straightforward, and the physical interpretation of the processes involved is simple. Our test-case shows that a profound enhancement of Faraday rotation can be achieved, thus potentially establishing a platform for non-reciprocal optical devices at relatively weak magnetic fields. While the study was carried on a simple test-case, we believe that our conclusions can be generalized and applied to other configurations that support dark-mode states.

\section{Additional content}
The Supporting Information includes detailed derivation of the polarizability expressions used throughout the article, in addition to a discussion regarding the response of the core and the shell separately, a detailed mapping of the resonances in the system, details regarding the application of the expressions to the screen problem and regarding the numerical simulations, and figures describing the effect of material loss. This material is available free of charge via the Internet at http://pubs.acs.org

\vspace*{0.5in}
M.~Meir and Y.~Mazor contributed equally to this work.

%\input{References_CoreShell_Mag_YMazor_MMeir_BZS_R1.tex}

%%%%%%%%%%%%%%%%%%%%%% SI %%%%%%%%%%%%%%%%%%%%%%
\section{Supporting Information}
\subsection{Introduction}

Since the structure under study is much smaller than the wavelength, we use Discrete Dipole Approximation (DDA) and polarizability theory to study its response to an impinging electromagnetic wave. The results of this study are used as initial exploration of the physical phenomena reported in the present work, and they are validated by exact full-wave simulations based on commercial numerical software.

The static polarizability $\bb{\alpha}_h$ is derived in Sec.~\ref{Intro} in addition to a discussion regarding the response of the core and the shell separately. The analysis below provides not only the net dipole response of the entire particle, but also the dipole densities $\bb{P}_{1,2}$ excited in the shell and the core regions, as well as the dipolar moment of each of these domains. In Sec.~\ref{StructRes} we demonstrate the structure resonances near the dark-mode and their variation due to a bias magnetic field. Details regarding the application to Faraday-rotating screen are discussed in Sec.~\ref{Screen}. The effect of material loss is discussed in Sec.~\ref{Loss}, and details pertaining to the full-wave numerical simulations are provided in Sec.~\ref{Numerical}.

\subsection{Formulation}
\label{Intro}

Under magnetization $\bb{B}_0=\unit{y}B_0$, $\bb{\chi}$ of a Drude-model metal is given by
\begin{equation}
\bb{\chi}=\frac{-\bar{\omega}^{-2}}{\bar{\omega}^2-\bar{\omega}_b^2}\!
 \left(
\begin{array}{ccc}
\bar{\chi}_{xx} & 0 & \bar{\chi}_{xz}\\
0 & \bar{\chi}_{yy} & 0\\
-\bar{\chi}_{xz} & 0& \bar{\chi}_{zz}\end{array}\right),\label{eq1si}
\end{equation}
with $\bar{\chi}_{xx}=\bar{\chi}_{zz}= \bar{\omega}^2$,
$\bar{\chi}_{yy}= \bar{\chi}_{xx}-\bar{\omega}_b^2$, and
$\bar{\chi}_{xz}= i\bar{\omega}\bar{\omega}_b$, and
where $\bar{\omega}=\omega/\omega_p$, $\bar{\omega}_b=\omega_b/\omega_p$, and where $\omega_p$ and $\omega_b=-q_eB_0/m_e$ are the plasma and cyclotron frequencies.
The validity of \Eq{eq1si} for nano-scale metal structures has been verified experimentally. The review paper \cite{R1} uses this formulation to discuss the magneto-optical effect in nanometer thin metal layers.
Furthermore, the work in \cite{R2} provides \emph{experimental} evidence for the validity of this formulation in nano-meter thin metal layers. The magnetic bias is parallel to the metal layers hence gyration plane of the electrons is normal to the surface. Metal layers of thickness down to 20 nm were tested; there is a nearly perfect fit between the measured data and the theoretical results.

The scalar electric potential $\Phi$ in non-isotropic media satisfies the generalized Laplace equation
$\nabla\cdot\bb{\epsilon}\nabla\Phi=0$. With $\bb{\chi}$ above we have $\epsilon_{xx}=\epsilon_{zz}$, and $\epsilon_{xz}=-\epsilon_{zx}$. Due to the latter the mixed derivative terms cancel. The equation then reads
\begin{equation}\label{eq2si}
\left[\epsilon_{xx}\partial_x^2+\epsilon_{yy}\partial_y^2+\epsilon_{xx}\partial_z^2\right]\Phi=0.
\end{equation}
If $\epsilon_{yy}\neq\epsilon_{xx}$ this equation is separable only in oblate or prolate spheroidal coordinates, while our geometry is of spherical symmetry. However, we note that in the regime of parameters relevant for the present work, the \emph{relative} difference between $\epsilon_{xx}$ and $\epsilon_{yy}$ is in the order of $10^{-6}$, and is negligible. We therefor approximate $\epsilon_{yy}\approx \epsilon_{xx}$ to render the Laplace operator above separable in the spherical system.

As noted in the main text, the shell (core) is made of Drude-metal with plasma frequency $\omega_{p1}$ ($\omega_{p2}$) and tensor electric susceptibility $\bb{\chi}_1$ ($\bb{\chi}_2$).
The outer shell radius and core radius are $a_1$ and $a_2$, respectively.
It is convenient to express the electric scalar potential in the entire space by
the three scalar potentials $\Phi_{e,1,2}$ for the external domain ($r>a_1$), shell ($a_2< r\leq a_1$) and core ($r\leq a_2$), respectively. We assume now that the particle is excited by $\unit{z}$-directed local electric field $E^L$; it is the field in the region, but in the absence of the particle.
The corresponding potential is $-E^L\,z=-E^L\,r\cos\theta$. Clearly, $\Phi_e$ should reduce to the latter when $r\gg a_1$, and $\Phi_{1,2}$ should remain bounded within their respective domains.
Up to the first order spherical harmonics (that amount to constant fields + dipole corrections)
$\Phi_{e,1,2}$ can be written in terms of the unknown polarizability components $\alpha_{h,zz},\alpha_{xz}=-\alpha_{zx}$ and six more unknowns, namely $A_1,B_1,C_1,D_1,A_2,C_2$,
\begin{subequations}
\begin{eqnarray}
\Phi_e &=& E^L\left(-r+\frac{\alpha_{h,zz}}{4\pi\epsilon_0 r^2}\right)\cos\theta\label{eq3asi}\\
       &+& \frac{E^L\alpha_{xz}}{4\pi\epsilon_0 r^2}\cos\phi\sin\theta\nonumber \\
       \nonumber\\
\Phi_1 &=& \left( A_1r+\frac{B_1}{r^2}\right)\cos\theta\label{eq3bsi}\\
       &+& \left( C_1r+\frac{D_1}{r^2}\right)\cos\phi\sin\theta \nonumber \\
       \nonumber\\
\Phi_2 &=& A_2r\cos\theta+C_2r\cos\phi\sin\theta.\label{eq3csi}
\end{eqnarray}
\end{subequations}
We emphasize that since $\Phi_e$ is the potential in the external domain, the terms $E^L\alpha_{h, zz}$ and $E^L\alpha_{h, zz}$ in \Eq{eq3asi} are the \emph{net} dipole moments in the $\unit{x}$ and $\unit{z}$ directions, respectively, due to the $\unit{z}$-directed local field $E^L$. Hence the unknown coefficients $\alpha_{h,zz},\alpha_{xz}=-\alpha_{zx}$ are indeed the respective polarizability components.

By imposing the continuity of the tangential $\bb{E}$ and normal $\bb{D}$ ($=\unit{r}\cdot\bb{D}$) across the core and shell boundaries, we derive a matrix equation for the eight unknowns. Here one needs to employ the susceptibilities in spherical coordinates, obtained by mapping the cartezian expressions of the form of \Eq{eq1si} via $\bb{\chi}_{1,2}\mapsto\mathcal{R}\bb{\chi}_{1,2}\mathcal{R}^{-1}$, where $\mathcal{R}$ is the cartezian-to-spherical transformation. The algebra is tedious but straightforward. The end result is the matrix equation
\begin{subequations}
\begin{equation}\label{eq4asi}
{\bf M}\bb{a}=\bb{F}
\end{equation}
with the right hand side vector
\begin{equation}\label{eq4bsi}
\bb{F}=(E^L,0,0,0,E^L,0,0,0)^T
\end{equation}
with the unknown vector $\bb{a}$
\begin{equation}\label{eq4csi}
\bb{a}=(B_e,D_e,A_1,B'_1,C_1,D'_1,A_2,C_2)^T
\end{equation}
and where we used $B_e=E^L\alpha_{h, zz}/(3V\epsilon_0)$,
$D_e=E^L\alpha_{xz}/(3V\epsilon_0)$, $B'_1=B_1/a_1^3$, and $D'_1=D_1/a_1^3$, with $V=4\pi a_1^3/3$ the total particle volume. $\bf M$ is the matrix
\begin{equation}\label{eq4dsi}
\left(
\begin{array}{cccccccc}
1     & 0    & -1              & -1               & 0              & 0                  & 0    & 0    \\
0     & 1    &  0              & 0                  & -1             & -  1               & 0    & 0    \\
0     & 0    &  1              & u                  & 0              & 0                  & -1   & 0     \\
0     & 0    &  0              & 0                  & 1              & u                & 0    & -1   \\
-2    & 0    &-\epsilon_{1,xx} &2\epsilon_{1,xx} & \chi_{1,xz}       & \chi_{1, xz}   & 0    & 0    \\
0     & -2   & -\chi_{1, xz}   & -\chi_{1, xz}  &-\epsilon_{1,xx} & 2\epsilon_{1,xx}& 0    & 0    \\
0     & 0    &-\epsilon_{1,xx} & 2u\epsilon_{1,xx} & \chi_{1, xz}  & u\chi_{1, xz}    & \epsilon_{2,xx}& -\chi_{2, xz}\\
0     & 0    & -\chi_{1,xz}    &-u\chi_{1,xz}    & -\epsilon_{1,xx} & 2u\epsilon_{1,xx}&\chi_{2,xz} &\epsilon_{2,xx}
\end{array}\right)
\end{equation}
\end{subequations}
with $u=(a_1/a_2)^3=V/V_2$ where $V$ is the particle volume and $V_2$ is the core volume.
This equation can be easily solved numerically. Then, the static net polarizability $\bb{\alpha}_h$ of the entire structure is readily obtained from $B_e,D_e$. The quasistatic polarizability $\bb{\alpha}$ is derived from $\bb{\alpha}_h$ via the incorporation of the radiation damping \cite{RadiativeCorrectionSI}
\begin{equation}
\bb{\alpha}^{-1}=\bb{\alpha}_h^{-1}-\frac{ik^3}{6\pi\epsilon_0}\bb{I}
\label{eq5si}
\end{equation}

\subsection{The shell and core responses}

Once the unknown coefficients are determined, the particle's polarizability matrix, the fields, and the internal polarization densities $\bb{P}_{1,2}=\epsilon_0\bb{\chi}_{1,2}\bb{E}_{1,2}$ in the shell and core are readily obtained.
The \emph{total} dipole moments $\bb{p}_{1,2}$ of the shell and core are of special interest to the study of the underlying physics. They are given by volume integration of the densities $\bb{P}_{1,2}$,
\begin{equation}
\bb{p}_1 = -\epsilon_0\int_{V_1}\bb{\chi}_1\nabla\Phi_1 dv
         = -\epsilon_0V_1\bb{\chi}_1
\left(
\begin{array}{c}
C_1\\
A_1
\end{array}\right)\label{eq6si}
\end{equation}
and similarly for $\bb{p}_2$,
\begin{equation}
\bb{p}_2 = -\epsilon_0\int_{V_2}\bb{\chi}_2\nabla\Phi_2 dv
         = -\epsilon_0V_2\bb{\chi}_2
\left(
\begin{array}{c}
C_2\\
A_2
\end{array}\right)\label{eq7si}
\end{equation}
where $V_{1,2}$ are the shell and core volumes.
In fact, \Eqs{eq6si}{eq7si} above provide explicit expressions for the \emph{separate} static polarizabilities $\bb{\alpha}_{h\, 1,2}$ of the shell and the core. If \Eqs{eq4asi}{eq4dsi} are solved for $E^L=1$V/m we may write
\begin{equation}\label{eq8si}
\bb{p}_{1,2}=\bb{\alpha}_{h\, 1,2}\bb{E}^L
\end{equation}
with
\begin{subequations}
\begin{eqnarray}
\alpha_{h1,xz} &=& -\epsilon_0V_1(\chi_{1,xx}C_1+\chi_{1,xz}A_1)\label{eq9asi}\\
\alpha_{h1,zz} &=& -\epsilon_0V_1(-\chi_{1,xz}C_1+\chi_{1,zz}A_1)\label{eq9bsi}\\
\alpha_{h2,xz} &=& -\epsilon_0V_2(\chi_{2,xx}C_2+\chi_{2,xz}A_2)\label{eq9csi}\\
\alpha_{h2,zz} &=& -\epsilon_0V_2(-\chi_{2,xz}C_2+\chi_{2,zz}A_2)\label{eq9dsi}
\end{eqnarray}
\end{subequations}
and with the symmetries $\alpha_{hn,xx}=\alpha_{hn,zz}$, $\alpha_{hn,zx}=-\alpha_{hn,xz}$ for $n=1,2$. These matrices are indeed hermitian since the $zz$ ($xz$) components have null imaginary (real) part.

\subsubsection{Radiative correction for the shell and core}

Since $\bb{p}_{1,2}$ are the total dipolar moment of the shell and the core, one obviously must have
\begin{equation}\label{eq10si}
\bb{p}=\bb{p}_1+\bb{p}_2\,\Leftrightarrow\, \bb{\alpha}=\bb{\alpha}_1+\bb{\alpha}_2
\end{equation}
This equality should hold for both the static and the quasistatic polarizabilities. While its validity in the static case is guaranteed via the exact solution of \Eqs{eq4asi}{eq4dsi}, the quasistatic expression is less transparent; one has to \emph{derive} the appropriate \emph{unique} radiation damping corrections for $\bb{p}_1$ and $\bb{p}_2$. These corrections are not obtained by a mere application of \Eq{eq10si} to $\bb{\alpha}_{1,2}$ since they mutually transfer power to the far field. To solve this problem we reconcile the DDA with the optical theorem by following essentially the same steps as in \cite{RadiativeCorrectionSI}.

The average power radiated to the far field by a single point dipole is well known. Its extension to the case of two dipoles $\bb{p}_{1,2}$ located at the origin is straightforward
%\begin{eqnarray}
%\mathcal{P}_r &=&     \frac{k^4c}{12\pi\epsilon_0}\label{eq11}\\
%           &\times& \left(\bb{p}_1^\dagger\bb{p}_1+\bb{p}_2^\dagger\bb{p}_2
%+\bb{p}_1^\dagger\bb{p}_2+\bb{p}_2^\dagger\bb{p}_1\right)\nonumber
%\end{eqnarray}
\begin{equation}\label{eq11si}
\mathcal{P}_r =\frac{k^4c}{12\pi\epsilon_0}
            \left(\bb{p}_1^\dagger\bb{p}_1+\bb{p}_2^\dagger\bb{p}_2
+\bb{p}_1^\dagger\bb{p}_2+\bb{p}_2^\dagger\bb{p}_1\right)
\end{equation}
where $^\dagger$ denotes transpose + complex conjugate. Likewise, the power transferred by the local field to the particle's currents is
\begin{equation}\label{eq12si}
\mathcal{P}_\ell=\frac{1}{2}\Re\int_V\bb{J}^*\cdot\bb{E}^Ldv=\frac{1}{2}\omega\Im\{\bb{E}^{L\dagger}\bb{p}\}.
\end{equation}
 Using \Eq{eq10si} in $\mathcal{P}_\ell$ and equating $\mathcal{P}_\ell=\mathcal{P}_r$ (in the absence of loss), we obtain
\begin{eqnarray}
& &\frac{k^3}{6\pi\epsilon_0}
\left[
\bb{p}_1^\dagger\bb{p}_1+\bb{p}_2^\dagger\bb{p}_2+
\bb{p}_1^\dagger\bb{\alpha}_2\bb{\alpha}_1^{-1}\bb{p}_1 +
\bb{p}_2^\dagger\bb{\alpha}_1\bb{\alpha}_2^{-1}\bb{p}_2\right]\nonumber\\
& & =
\Im\left(\bb{E}^{L\dagger}\bb{\alpha}_1\bb{E}^L+\bb{E}^{L\dagger}\bb{\alpha}_2\bb{E}^L\right)\label{eq13si}
\end{eqnarray}
where we used $\bb{p}^\dagger_{1,2}=\bb{E}^{L\dagger}\bb{\alpha}^\dagger_{1,2}$, $\bb{p}_1^\dagger\bb{p}_2=\bb{p}_1^\dagger\bb{\alpha}_2\bb{E}^L=\bb{p}_1^\dagger\bb{\alpha}_2\bb{\alpha}_1^{-1}\bb{p}_1$ and similarly
$\bb{p}_2^\dagger\bb{p}_1=\bb{p}_2^\dagger\bb{\alpha}_1\bb{\alpha}_2^{-1}\bb{p}_2$. In the last line we use now
\begin{equation}\label{eq14si}
\bb{E}^{L\dagger}\bb{\alpha}_n\bb{E}^L=
\bb{E}^{L\dagger}\bb{\alpha}^\dagger_n\bb{\alpha}^{\dagger\, -1}_n\bb{\alpha}_n\bb{E}^L=
\bb{p}^\dagger_n\bb{\alpha}^{\dagger\, -1}_n\bb{p}_n
\end{equation}
with $n=1,2$. Then \Eq{eq13si} can be written as
\begin{equation}\label{eq15si}
\Im\left(\bb{p}^\dagger_1   \bb{\alpha}_{h1}^{-1}     \bb{p}_1+
\bb{p}^\dagger_2 \bb{\alpha}_{h2}^{-1} \bb{p}_2\right)=0
\end{equation}
where $\bb{\alpha}_{h n}^{-1}$ are the matrices
\begin{subequations}
\begin{eqnarray}
\bb{\alpha}_{h 1}^{-1}&=&
        \bb{\alpha}_{1}^{\dagger\,-1}-iR({\bf I}+\bb{\alpha}_2\bb{\alpha}_1^{-1})\label{eq16asi}\\
\bb{\alpha}_{h 2}^{-1}&=&
        \bb{\alpha}_{2}^{\dagger\,-1}-iR({\bf I}+\bb{\alpha}_1\bb{\alpha}_2^{-1})\label{eq16bsi}
\end{eqnarray}
\end{subequations}
where $R=k^3/(6\pi\epsilon_0)$. Note that \Eq{eq15si} is in fact a sum of two quadratic forms. Hence, a \emph{sufficient} condition satisfying this equation is that these matrices be hermitian. Using again \Eq{eq10si}, the last two equations can be rewritten as ($n=1,2$)
\begin{equation}
\bb{\alpha}_{h n}^{-1}=
        \bb{\alpha}_{n}^{\dagger\,-1}-iR\bb{\alpha}\bb{\alpha}_n^{-1}\label{eq17si}
\end{equation}
The hermitian matrices $\bb{\alpha}_{h n}$ are nothing but the corresponding static polarizabilities \cite{RadiativeCorrectionSI}. Furthermore, borrowing from the structure of \Eq{eq5si}, one may express the inverse dynamic polarizabilities $\bb{\alpha}^{-1}_{1,2}$ as
\begin{equation}\label{eq18si}
\bb{\alpha}_n^{-1}=\bb{\alpha}_{h n}^{-1}-i{\bf R}_n,\qquad n=1,2
\end{equation}
where ${\bf R}_n$ is the radiative correction, yet to be determined. An explicit equation can now be obtained for ${\bf R}_n$,
\begin{equation}\label{eq19si}
{\bf R}_n=({\bf I}+iR\bb{\alpha})^{-1}R\bb{\alpha}\bb{\alpha}_{h n}^{-1}.
\end{equation}
Note that the radiation loss associated with each of the particle components (${\bf R}_{1,2}$ for the shell and core), depends on $\bb{\alpha}$ - the polarizability of the entire structure.

\subsection{Structure resonances}\label{StructRes}

\begin{figure}[h]
\begin{center}
\noindent
\hspace*{-0.27in}
  \includegraphics[width=8cm]{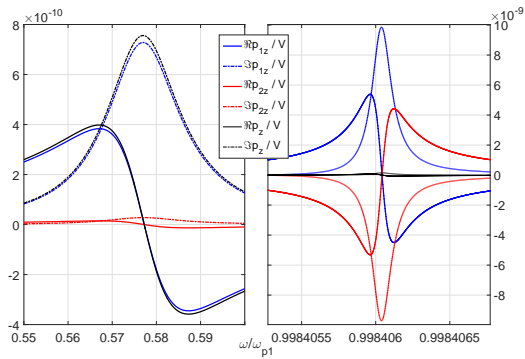}
  \caption{The lower and upper resonances.}
  \label{fig1si}
\end{center}
\end{figure}

With no magnetization ($B_0=0$), the structure under study possesses two quasi-static resonances. Figure \ref{fig1si} shows the dipole response for $a_1/a_2=3$ and $\omega_{p1}/\omega_{p2}=1.005$ over the frequency ranges of the lower and upper resonances.
It is clearly seen that in the frequency domain of the lower resonance $\bb{p}_1$ and $\bb{p}_2$ are in phase and interfere constructively. Furthermore - the polarization \emph{densities} in the core and shell are nearly the same; the core dipole $\bb{p}_2$ is about 26 times weaker because the core volume is 26 times smaller than the shell volume ($(a_1/a_2)^3-1=26$). Therefore, the dynamics of the core-shell particle in the lower resonance is very similar to that of a conventional spherical particle. However, in the upper domain $\bb{p}_1$ and $\bb{p}_2$ resonate intensely (their magnitude is 10-times larger than in the lower domain), but they are of opposite signs and nearly cancel over the entire domain. The net response possesses a weaker magnitude with a very narrow line-shape (see also Fig.~3 in the main text).

\begin{figure}[h]
\begin{center}
\noindent
\hspace*{-0.27in}
  \includegraphics[width=7.5cm]{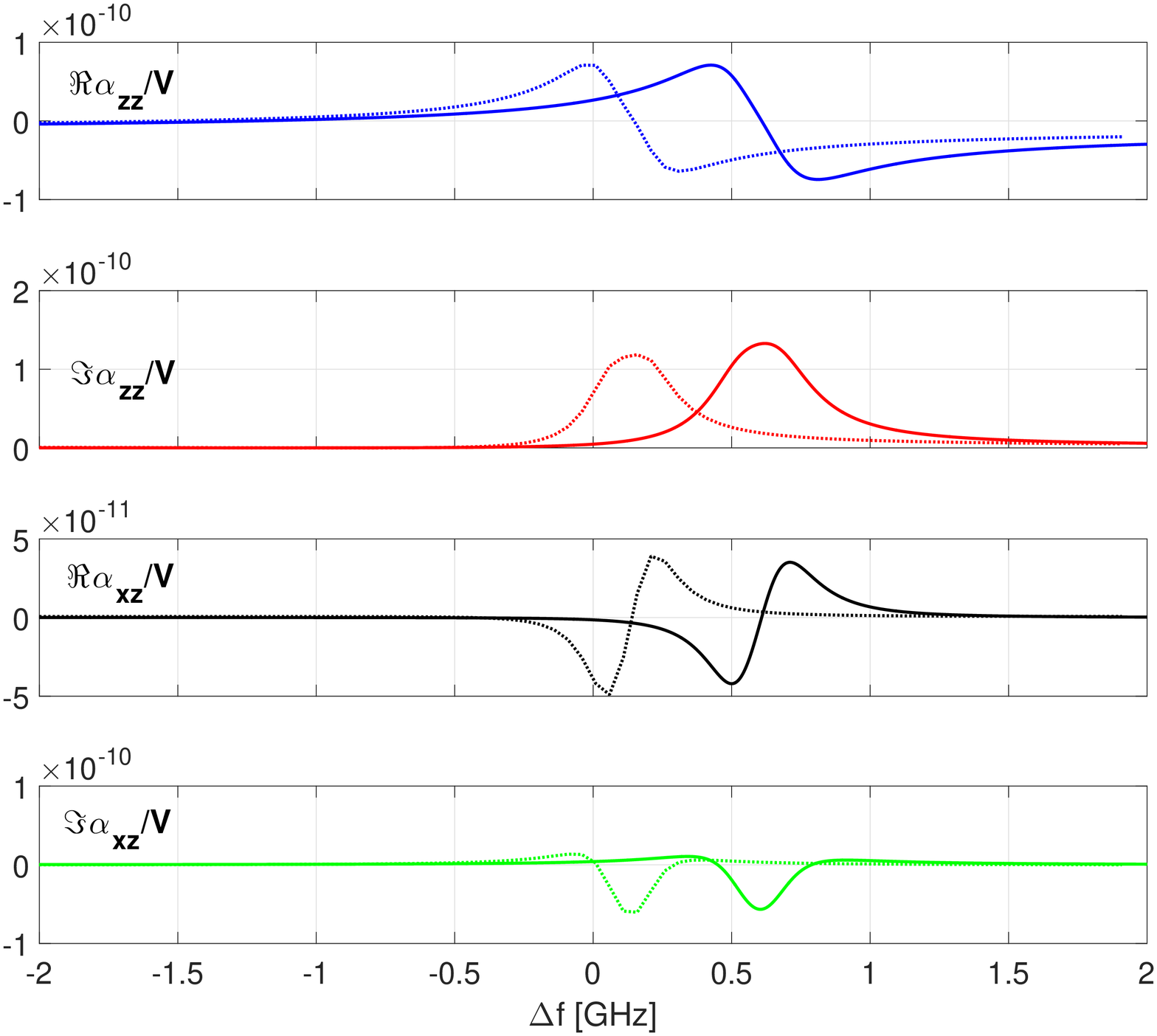}
  \caption{Core-shell particle polarizability at the upper resonance under magnetization of 1.5Tesla. $\Delta f=0$ corresponds to the upper resonance frequency of the unmagnetized core-shell structure. Solid lines show the results of the polarizability theory. Dashed lines show the full-wave numerical simulations using the $\mbox{CST}^\copyright$.}
  \label{fig2si}
\end{center}
\end{figure}

The results obtained by the theory developed here in \Eqs{eq4asi}{eq5si} are compared to full-wave simulations using the $\mbox{CST}^\copyright$ commercial software. Comparison of the polarizability with no magnetization is included in Fig.~3 in the main text. Figure \ref{fig2si} here shows a comparison under magnetization of 1.5Tesla, with the parameters used in the text. The comparison is to the results shown in Fig.~5 in the main text. It is seen that the results agree very well.

The Faraday rotation of the net dipolar response is fully characterized by the dipole tilt and ellipticity angles $\psi,\theta_e$, as discussed in the main text. Figure \ref{fig3si} below shows $\psi,\theta_e$ for a conventional spherical particle made of \emph{lossless} Ag, under a bias magnetic field of $B_0=1.5$T and the same diameter as the outer dimension of our core-shell particle. Due to $B_0$ the particle's single resonance splits to two different resonance frequencies. The tilt angle is maximized at these frequencies. However, it is seen that the maximal achievable tilt is about $0.06^\circ$. This is to contrast with the core-shell geometry where rotation angles $>20^\circ$ are obtained - see discussion and Fig.~6 in the main text.

\begin{figure}[h]
\begin{center}
\noindent
\hspace*{-0.27in}
  \includegraphics[width=7.5cm]{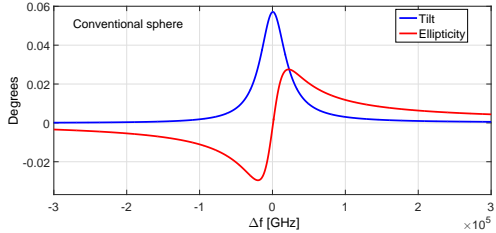}
  \caption{Dipole tilt and ellipticity angles for a \emph{lossless} Ag particle of conventional structure. The bias magnetic field is $B_0=1.5$T. $\Delta f=0$ corresponds to the resonant frequency of the unmagnetized plasmonic sphere $\omega_p/(2\pi\sqrt{3})$. The maximal achievable rotation is a fraction of a degree.}
  \label{fig3si}
\end{center}
\end{figure}

\subsection{Application to Faraday rotating screen}\label{Screen}

\begin{figure}[h]
\begin{center}
\noindent
\hspace*{-0.27in}
  \includegraphics[width=8.0cm]{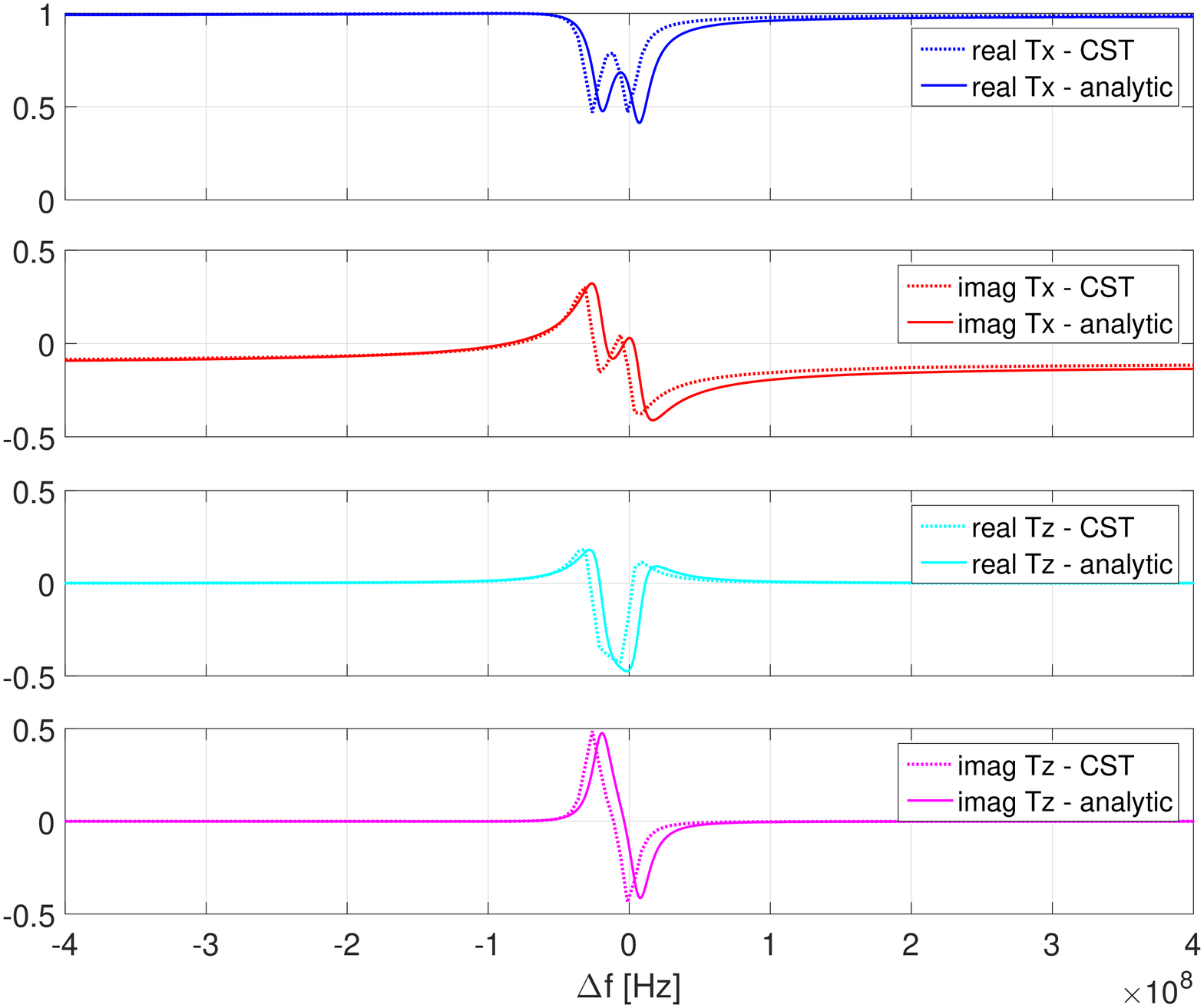}
  \caption{Transmission through the Faraday rotating screen described in the Application section of the main text. Solid lines show the results of the full-wave simulation with the CST$^\copyright$ package. Dashed lines show the results obtained using the semi-analytical method discussed in the main text. $\Delta f=0$ corresponds to the frequency $f=2067.3371211$THz.}
  \label{fig4si}
\end{center}
\end{figure}

We use the core-shell particle co construct a non-reciprocal metasurface for Faraday rotation application. The geometry and its parameters are shown and discussed in the Application section of the main text. Here, we compare the transmission coefficient results obtained by using the semi-analytical method discussed in the main text, to those of full-wave numerical simulations using the CST$^\copyright$. The comparison is shown in Fig.~\ref{fig4si} for a magnetic bias strength is 1.5Tesla (i.e. the comparison is done along a horizontal line of Fig.~8 of the main text). It is seen that the results agree well.

\subsection{The effect of material loss}\label{Loss}

Like in many other plasmonic applications and non-reciprocal effects, the presence of loss reduces system performances. For brevity, the present discussion is limited to the effect of loss on the performance of the ``end-product'' -  the Faraday rotating screen. The effect of loss on the dipole rotation in a single particle is essentially the same.
We characterize material loss by the parameter $\sigma=(\tau\omega_{p1})^{-1}$ where $\tau$ is the dissipation time constant.

\begin{figure}[h]
\begin{center}
\noindent
\hspace*{-0.27in}
  \includegraphics[width=7.0cm]{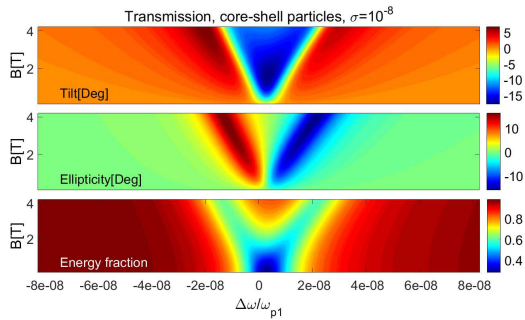}
  \caption{The properties of transmission through the Faraday rotating screen described in the Application section of the main text, with $\sigma=10^{-8}$. }
  \label{fig5si}
\end{center}
\end{figure}

\begin{figure}[h]
\begin{center}
\noindent
\hspace*{-0.27in}
  \includegraphics[width=7.0cm]{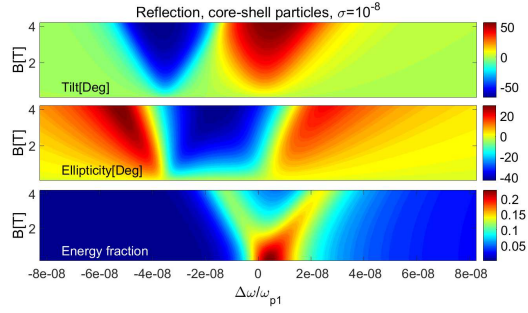}
  \caption{The same as Fig.~\ref{fig5si}, but for reflection.}
  \label{fig6si}
\end{center}
\end{figure}

Figures \ref{fig5si} and \ref{fig6si} show the transmission and reflection properties of a screen with parameters similar to the lossless screen discussed in the main text and with loss parameter $\sigma=10^{-8}$. It is seen that at $B_0=1.5$T one may still get tilt of $15-20^\circ$ and low ellipticity. When $\sigma$ is further increased the corresponding tilt reduces.

\begin{figure}[h]
\begin{center}
\noindent
\hspace*{-0.27in}
  \includegraphics[width=7.0cm]{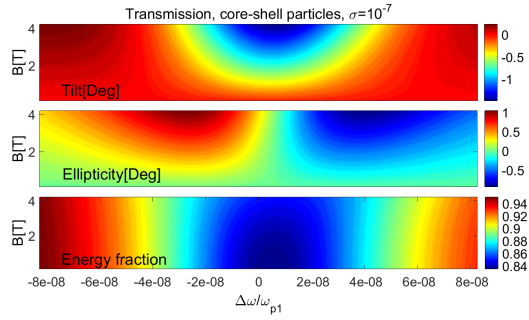}
  \caption{The properties of transmission through the Faraday rotating screen described in the Application section of the main text, with $\sigma=10^{-7}$. }
  \label{fig7si}
\end{center}
\end{figure}
\begin{figure}[h]
\begin{center}
\noindent
\hspace*{-0.27in}
  \includegraphics[width=7.0cm]{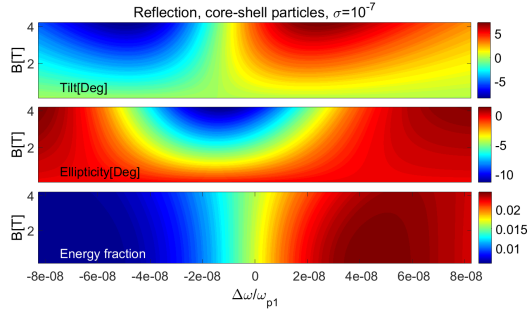}
  \caption{The same as Fig.~\ref{fig7si}, but for reflection.}
  \label{fig8si}
\end{center}
\end{figure}

Figures \ref{fig7si} and \ref{fig8si} show the same properties, but with $\sigma=10^{-7}$. For magnetic bias of 1.5T the tilt angle is in the order of $0.5^\circ$, and it increases to about $1.5^\circ$ for $B_0=4$T. We emphasize that although there is a significant reduction of the Faraday rotation effect in the presence of this loss, both cases ($\sigma=10^{-8}$ and $\sigma=10^{-7}$) provide rotations that are much larger than the rotation obtained in a single \emph{lossless} particle - see for example Fig.~\ref{fig3si} above.


\begin{figure}[h]
\begin{center}
\noindent
\hspace*{-0.27in}
  \includegraphics[width=7.0cm]{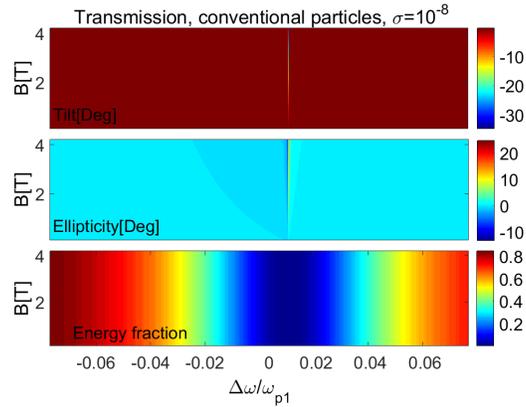}
  \caption{The properties of transmission through a Faraday rotating screen that consists of conventional plasmonic particles, with $\sigma=10^{-8}$. }
  \label{fig9si}
\end{center}
\end{figure}

\begin{figure}[h]
\begin{center}
\noindent
\hspace*{-0.27in}
  \includegraphics[width=7.0cm]{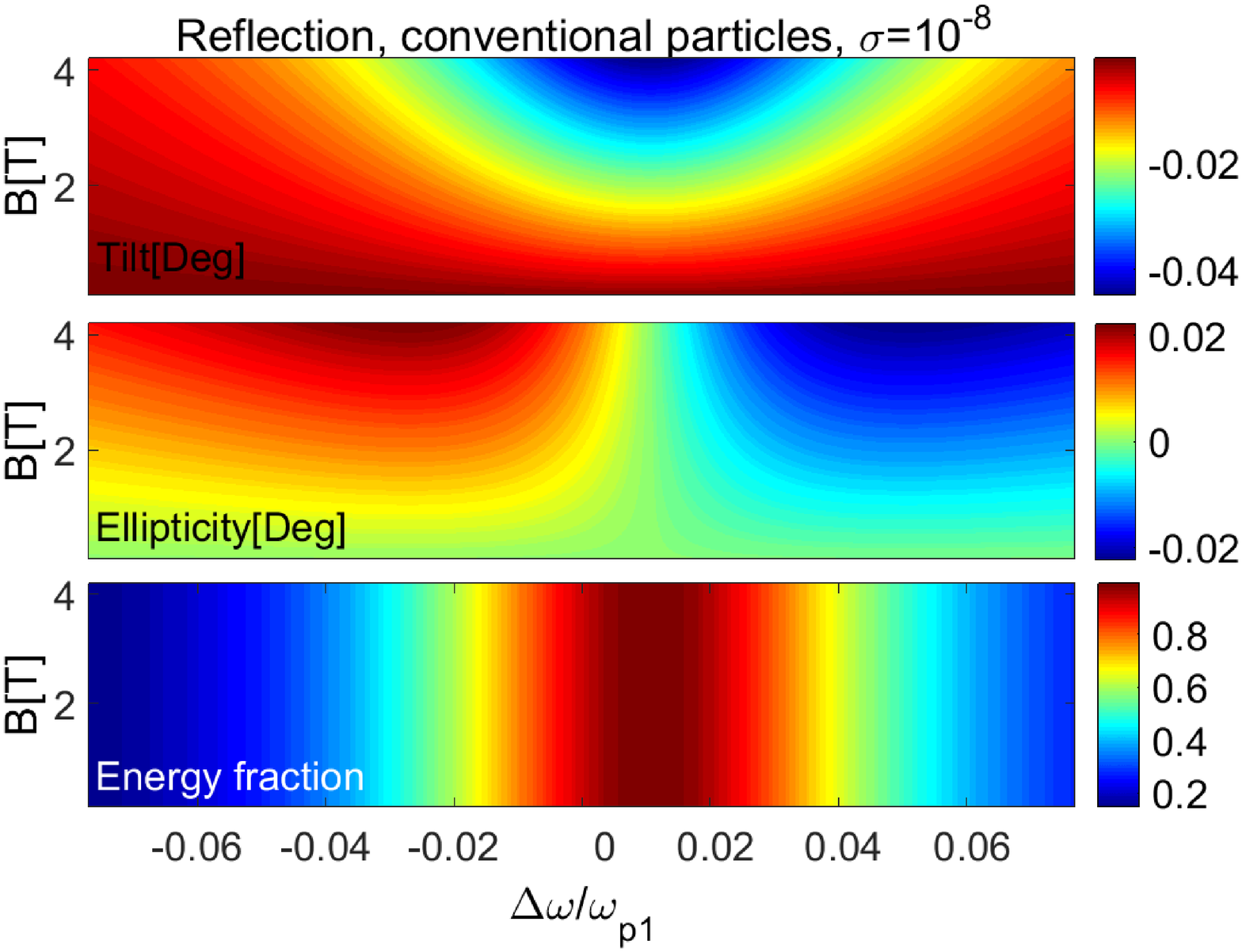}
  \caption{The same as Fig.~\ref{fig9si}, but for reflection.}
  \label{fig10si}
\end{center}
\end{figure}

\begin{figure}[h]
\begin{center}
\noindent
\hspace*{-0.27in}
  \includegraphics[width=7.0cm]{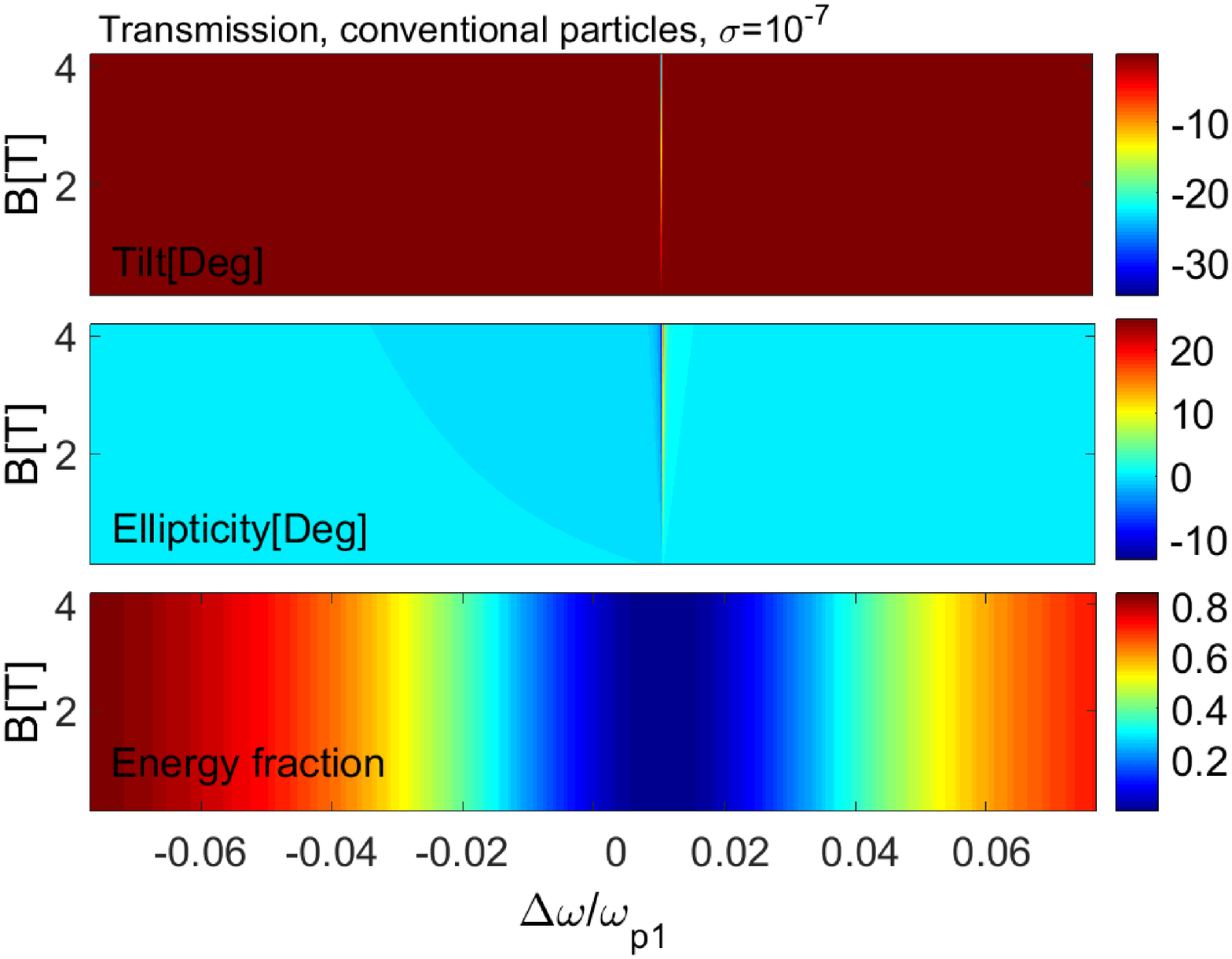}
  \caption{The properties of transmission through a Faraday rotating screen that consists of conventional plasmonic particles, with $\sigma=10^{-7}$. }
  \label{fig11si}
\end{center}
\end{figure}

\begin{figure}[h]
\begin{center}
\noindent
\hspace*{-0.27in}
  \includegraphics[width=7.0cm]{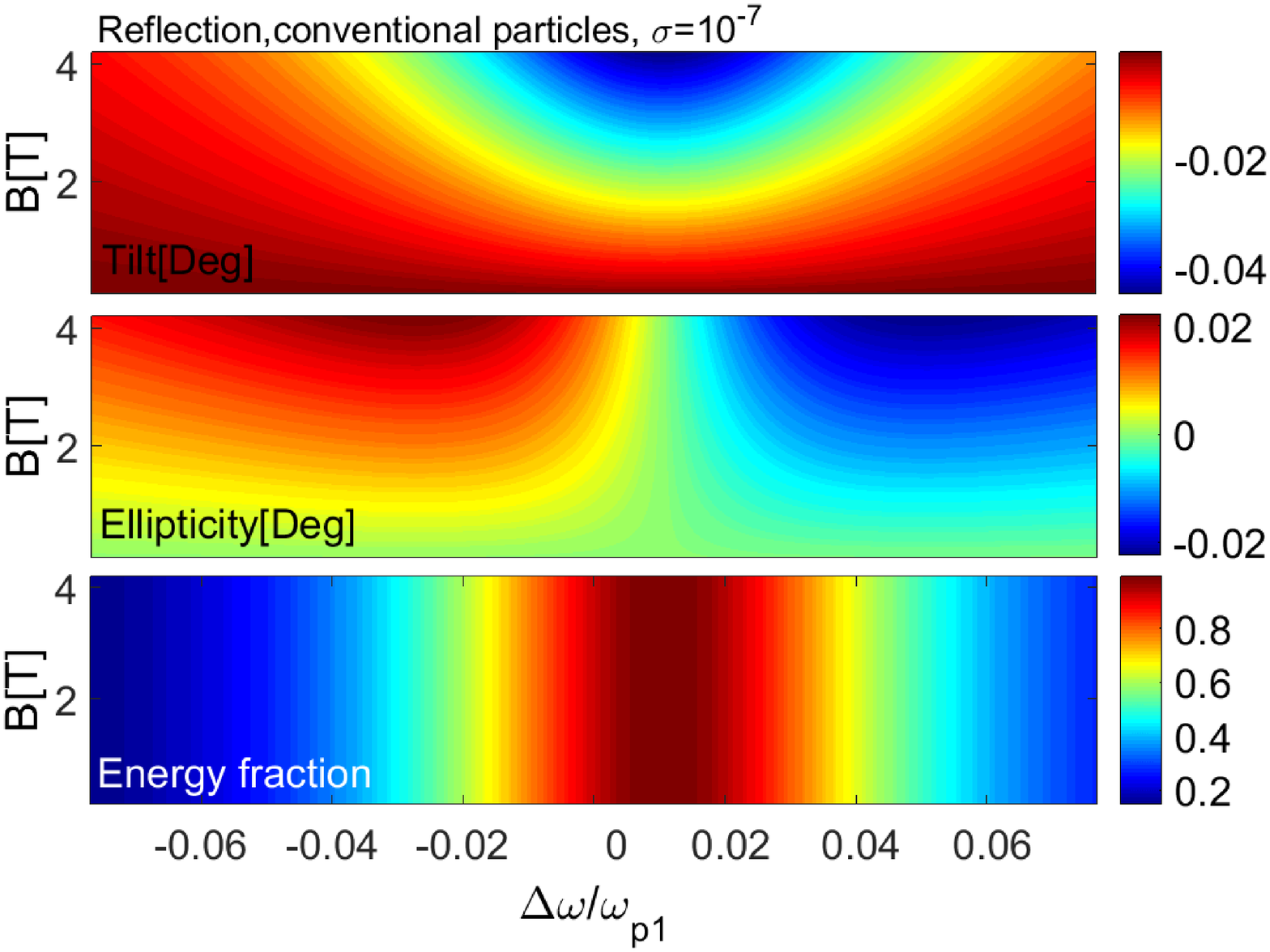}
  \caption{The same as Fig.~\ref{fig11si}, but for reflection.}
  \label{fig12si}
\end{center}
\end{figure}

For comparison, Figs.~\ref{fig9si}-\ref{fig10si} show the rotation properties of a similar screen that consists of conventional spherical particles of the same size, with loss $\sigma=10^{-8}$. Clearly, at resonance the screen transmission vanishes, hence the narrow vertical region of high tilt seen at the tilt and ellipticity panes of Fig.~\ref{fig9si} corresponds to the case of zero transmission has no physical significance. Figures~\ref{fig11si}-\ref{fig12si} show the same, but with $\sigma=10^{-7}$. Clearly, the core-shell structure outperforms the conventional particles structure by orders of magnitude.




\subsection{Numerical simulations - details}\label{Numerical}
Throughout this paper, CST software was used to perform numerical simulations.

For simulating a single particle we have constructed the core-shell geometry and used the drude-model option for material properties. The structure was discretized to ~260000 tetrahedral cells, and adaptive meshing was used to account for delicate field dynamics in certain domains. Stopping condition was defined by relative residual of $10^{-4}$. The problem was excited by a plane wave at each of the simulated frequencies.

For simulating an infinite 2D array we used the same geometrical and material model but with periodic boundary conditions. The unit cell was discretized to ~300000 tetrahedral cells and adaptive mesh was used. Stopping condition was the same - $10^{-4}$ for the relative residual. The problem was excited by a plane wave at each of the simulated frequencies.

% that's all folks

\begin{thebibliography}{1}

\bibitem{Stockman}
P.~Nordlander, C.~Oubre, E.~Prodan, K.~Li, and M.~I.~Stockman,
``Plasmon Hybridization in Nanoparticle Dimers,''
\emph{Nano Letters} {\bf 4}(5), 899-903 (2004).

\bibitem{BoundStatesPhotonics}
D.~C.~Marinica, A.~G.~Borisov, and S.~V.~Shabanov,
``Bound States in the Continuum in Photonics,''
\emph{Phys.~Rev.~Lett.} {\bf 100}, 183902 (2008).

\bibitem{BrightDark}
M.~W.~Chu, V.~Myroshnychenko, C.~H.~Chen, J.~P.~Deng, C.~Y.~Mou, and F.~J.~Garcia de Abajo,
``Probing Bright and Dark Surface-Plasmon Modes in Individual and Coupled Noble Metal Nanoparticles Using and Electron Beam,''
\emph{Nano Letters} {\bf 9}(1), 399-404 (2009).

\bibitem{Maier1}
S.~A.~Maier,
``The benefits of darkness,''
\emph{Nature Materials}, {\bf 8}, 699 (2009).

\bibitem{DMExcitationLocalizedEmitter}
M.~Liu, T.~L.~Lee, S.~K.~Gray, P.~G.~Sionnest, and M.~Pelton,
``Excitation of Dark Plasmons in Metal Nanoparticles by a Localized Emitter,''
\emph{Phys.~Rev.~Lett.} {\bf 102}, 107401 (2009)

\bibitem{DarkPlasmonics}
D.~E.~Gomez, Z.~Q.~Teo, M.~Altissimo, T.~J.~Davis, S.~Earl, and A.~Roberts,
``The Dark Side of Plasmonics,''
\emph{Nano Letters} {\bf 13}, 3722-3728 (2013).

\bibitem{DarkAlu}
F.~Montecone and A. Alu,
``Embedded Photonic Eigenvalues in 3D Nanostructures,''
\emph{Phys.~Rev.~Lett.} {\bf 112}, 213903 (2014).

\bibitem{FanoHalas}
B.~Luk'yanchuk, N.~I.~Zheludev, S.~A.~Maier, N.~J.~Halas, P.~Nordlander, H.~Giessen, and C.~T.~Chong,
``The Fano resonance in plasmonic nanostructures and metamaterials,''
\emph{Nature Materials} {\bf 9} 707 (2010).

%%%%%%% Magnetoplasmonics

\bibitem{Fan}
Z.~Yu, G.~Veronis, Z.~Wang, and S.~Fan,
``One-Way Electromagnetic Waveguide Formed at the Interface between a Plasmonic Metal
under a Static Magnetic Field and a Photonic Crystal,''
\emph{Phys. Rev. Lett.} {\bf 100}, 023902 (2008).

\bibitem{HadadSteinberg}
Y.~Hadad and Ben Z.~Steinberg,
``Magnetized Spiral Chains of Plasmonic Ellipsoids for One-Way Optical Waveguides,''
\emph{Phys. Rev. Lett.} {\bf 105}, 233904 (2010).

\bibitem{MazorSteinbergFlat}
Y.~Mazor and Ben Z.~Steinberg,
``Longitudinal chirality, enhanced nonreciprocity, and nanoscale planar one-way plasmonic guiding,''
\emph{Phys.~Rev.~B} {\bf 86}, 045120 (2012).

\bibitem{OneWayStrips}
Y.~Mazor, H.~Hadad, and Ben Z.~Steinberg,
``Planar one-way guiding in periodic particle arrays with asymmetric unit cell and general group-symmetry considerations,''
\emph{Phys. Rev. B} {\bf 92}, 125129 (2015).

\bibitem{Wire}
A.~Davoyan and N.~Engheta,
``Electrically controlled one-way photon flow in plasmonic nanostructures,''  % The current carrying wire one-way
\emph{Nature Communications} 6:5250 doi: 10.1038/ncomms6250 (2014).

\bibitem{Circulator}
A.~Davoyan and N.~Engheta,
``Nonreciprocal rotating power flow within plasmonic nanostructures,''
\emph{Phys. Rev. Lett.} {\bf 111}, 047401 (2012).



%%%%%%%%%%%%%%%%%%%%%%%%%%%%%%

\bibitem{GoldCoatedIronOxide}
P.~K.Jain, Y.~Xiao, R.~Walsworth, and A.~E.~Cohen,
``Surface plasmon resonance enhanced magneto-optics (SuPREMO): Faraday Rotation Enhancement in Gold-Coated Iron Oxide Nanocrystals,''
\emph{Nano Letters} {\bf 9}(4), 1644-1650 (2009).

\bibitem{Graphene1}
I.~Crassee, J.~Levallois, A.~L.~Walter, M.~Ostler, A.~Bostwick,
E.~Rotenberg, T.~Seyller, D.~van der Marel, A.~B.~Kuzmenko,
``Giant Faraday Rotation in Single- and Multilayer Graphene,''
\emph{Nat.~Phys.} {\bf 7}, 48-51 (2010).

\bibitem{GrapheneACS}
Y.~Hadad, A.~Davoyan, N.~Engheta, and Ben Z. Steinberg,
``Extreme and Quantized Magneto-optics with Graphene Meta-atoms and Metasurfaces,''
\emph{ACS Photonics} {\bf 1}, 1068-1073 (2014).

%%%%%%%% Loss %%%%%%%%%

\bibitem{Khurgin_Elusive} %[1]
J.~B.~Khurgin and G.~Sun,
``In search of the elusive losless metal''
\emph{Applied Physics Letters} {\bf 96}, 181102 (2010)

\bibitem{PRX}   %[2]
J.~Kim, G.~V.~Naik, A.~V.~Gavrilenko, K.~Dondapati, V.~I.~Gavrilenko,
S.~M.~Prokes, O.~J.~Glembocki, V.~M.~Shalaev, and A.~Boltasseva,
``Optical Properties of Gallium-Doped Zinc Oxide--Low-Loss Plasmonic Material:
First-Principles Theory and Experiment,''
\emph{Physical Review X} {\bf 3}, 041037 (2013).

\bibitem{Naik2} %[3]
G.~V.~Naik, V.~M.~Shalaev, and A.~Boltasseva,
``Alternative Plasmonic Materials: Beyond Gold and Silver''
\emph{Advanced Materials} {\bf 25}, 3264-3294 (2013).

\bibitem{MaradudinBook}   %[4]
A. A. Maradudin, J. R. Sambles, and W. L. Barnes, editors, \emph{Modern Plasmonics}, Elsevier 2014 (Chapt. 6)

\bibitem{PendrySpoof}   %[5]
J.~B.~Pendry, L.~Martin-Moreno, and F.~J.~Garcia-Vidal,
``Mimicking Surface Plasmons with Structured Surfaces''
\emph{Science} {\bf 305}, 847-848 (2004)

\bibitem{SpoofExperimental} %[6]
A.~P.~Hibbins, B.~R.~Evans, J.~R.~Sambles,
``Experimental Verification of Designer Surface Plasmons,''
\emph{Science} {\bf 308}, 670-672 (2005)

\bibitem{SpoofNonRecip}   %[7]
A.~B.~Khanikaev, S.~H.~Mousavi, G.~Shvets, and Y.~S.~Kivshar,
``One-Way Extraordinary Optical Transmission and Nonreciprocal Spoof Plasmons,''
\emph{Phys.~Rev.~Lett.} {\bf 105}, 126804 (2010)

\bibitem{Active_Book} % [8]
A.~V.~Zayats and S.~A.~Maier editors, \emph{Active Plasmonics and Tuneable Plasmonic Metamaterials}, Willey 2013 (Chapt. 1)

\bibitem{Capolino1} % [9]
S.~Campione, M.~Albani, and F.~Capolino,
``Complex modes and near-zero permittivity in 3D arrays of plasmonic nanoshells: loss compensation using gain,''
\emph{Optical Materials Express} {\bf 1}(6), 1077-1089 (2011)

\bibitem{Capolino2}   %[10]
V.~Pustovit, F.~Capolino, and A.~Aradian,
``Cooperative plasmon-mediated effects and loss compensation by gain dyes near a metal nanoparticle,''
\emph{J.~Opt.~Soc.~Am.~B} {\bf 32}(2), 188-193 (2015)

%%%%%%%%%%%%%%%%%%%%%%%%%%%%%%%%%

\bibitem{SihvolaBook}
A.~Sihvola, \emph{Electromagnetic mixing formulas and applications}, Electromagnetic Waves Series (IEE, London, 1999).

%%%%%%%%%%%% Core shell

\bibitem{AluEngheta1}
A.~Alu and N.~Engheta,
``Polarizabilities and effective parameters for collections of spherical
nanoparticles formed by pairs of concentric double-negative,
single-negative, and/or double-positive metamaterial layers,''
\emph{J. Appl. Phys.} {\bf 97}, 094310 (2005).

\bibitem{AluEngheta2}
A.~Alu and N. Engheta,
``Theory of linear chains of metamaterial/plasmonic particles as subdiffraction optical nanotransmission lines,''
\emph{Phys.~Rev. B} {\bf 74}, 205436 (2006).

\bibitem{Cloaks}
C.~Argyropoulos, P.~Y.~Chen, F.~Monticone, G.~D'Aguanno, and A.~Alu`,
``Nonlinear Plasmonic Cloaks to Realize Giant All-Optical Scattering Switching,''
\emph{Phys.~Rev.~Lett.} {\bf 108}, 263905 (2012).

\bibitem{TretyakovCoreShell}
Y.~Ra'di, V.~S.~Asadchy, S.~U.~Kosulnikov, M.~M.~Omelyanovich, D.~Morits,
A.~V. Osipov, C.~R.~Simovski, and S.~A.~Tretyakov,
``Full Light Absorption in Single Arrays of Spherical Nanoparticles,''
\emph{ACS Photonics} {\bf 2}, 653-660 (2015).

%%%%%%%%%%%%
% Plasma frequencies
%%%%%%%

\bibitem{PFreq1}
A. D. Rakic , A. B. Djurisic , J. M. Elazar, and M. L. Majewski,
``Optical properties of metallic films
for vertical-cavity optoelectronic devices,''
\emph{Applied Optics} {\bf 37}(22), 5271-5283 (1998).


\bibitem{PFreq2}
M. A. Ordal, Robert J. Bell, R. W. Alexander, Jr, L. L. Long, and M. R. Querry,
``Optical properties of fourteen metals in the infrared and
far infrared: Al, Co, Cu, Au, Fe, Pb, Mo, Ni, Pd,
Pt, Ag, Ti, V, and W.''
\emph{Applied Optics} {\bf 24}(24), 4493-4498 (1985).

%%%%%%%%%%%%%

\bibitem{RadiativeCorrection}
S.~Albaladejo,~R. Gomez-Medina,~L.~S.~Froufe-Perez, H. Marinchio, R.~Carminati, J.~F.~Torrado, G.~Armelles,
A.~Garcia-Martin, and J.~J.~Saenz1,
``Radiative corrections to the polarizability tensor of an electrically small anisotropic dielectric particle,''
\emph{Optics Express} {\bf 18}(4), 3556 (2010).

\bibitem{TretyakovBook}
S.~Tretyakov \emph{Analytical modeling in applied electromagnetics}, Artech House (2003).

\end{thebibliography}

\begin{thebibliography}{1}

\bibitem{R1}
Vasily V.~Temnov, ``Ultrafast acousto-magneto-plasmonics,'' \emph{Nature Photonics}, {\bf 6}, 728 Nov.~2012 (DOI:10.1038/NPHOTON.2012.220).

\bibitem{R2}
Haefner, P., Luck, E., and Mohler, E. ``Magnetooptical properties of surface plasma waves on copper, silver, gold, and aluminum'' \emph{Phys. Stat. Sol.} {\bf 185}, 289-299 (1994).

\bibitem{RadiativeCorrectionSI}
S.~Albaladejo,~R. Gomez-Medina,~L.~S.~Froufe-Perez, H. Marinchio, R.~Carminati, J.~F.~Torrado, G.~Armelles,
A.~Garcia-Martin, and J.~J.~Saenz1,
``Radiative corrections to the polarizability tensor of an electrically small anisotropic dielectric particle,''
\emph{Optics Express} {\bf 18}(4), 3556 (2010).

\end{thebibliography}
\end{document}